\documentclass[aps,floatfix,twocolumn,showpacs,showkeys,preprintnumbers,amsmath]{revtex4}
\usepackage{amssymb}
\usepackage[T1]{fontenc}
\usepackage[latin1]{inputenc}
\usepackage{graphicx}
\usepackage{dcolumn}
\usepackage{bm}

\newcommand{\be}{\begin{eqnarray}}
\newcommand{\ee}{\end{eqnarray}}

\def\k{{\rm k}}
\def\q{{\rm q}}

\def\e{{\rm e}}

\begin{document}

\title{Microscopic calculation of neutrino mean free path inside hot neutron matter}
\author{Jérôme Margueron}
\affiliation{GANIL CEA/DSM - CNRS/IN2P3 BP 5027 F-14076 Caen CEDEX 5, France}
\author{Isaac Vida\~na}
\affiliation{Dipartimento di Fisica ``Enrico Fermi'', Universit\`a di Pisa and \\ INFN Sezione 
di 
Pisa, Via Buonarroti 2,I-56127 Pisa, Italy}
\author{Ignazio Bombaci}
\affiliation{Dipartimento di Fisica ``Enrico Fermi'', Universit\`a di Pisa and \\ INFN Sezione 
di 
Pisa, Via Buonarroti 2,I-56127 Pisa, Italy}

\date{\today}

\begin{abstract}
We calculate the neutrino mean free path and the Equation
of State of pure neutron matter at finite temperature within a 
selfconsistent scheme based on the Brueckner--Hartree--Fock approximation.
We employ the nucleon-nucleon part of the recent realistic baryon-baryon interaction
(model NSC97e) constructed by the Nijmegen group. The temperatures
considered range from 10 to 80 MeV. We report on the calculation of the mean field, 
the residual interaction and the neutrino mean free path including short and long range 
correlations given by the Brueckner--Hartree--Fock plus Random Phase Approximation (BHF+RPA) framework. 
This is the first fully consistent calculation in hot neutron matter dedicated to neutrino mean free path. 
We compare systematically our results to those obtain with the D1P Gogny effective interaction, which is 
independent of the temperature. The main differences between the present calculation and those with nuclear
effective interactions come from the RPA corrections to BHF (a factor of about 8) while the temperature 
%effects 
lack of consistency
accounts for a factor of about 2.
\end{abstract}

%\preprint{PISA-XXX-TH}



\maketitle

%%%%%%%%%%%%%%%%%%%%%%%%%%%%%%%%%%%%%%%%%%%%%%%%%%%%%%%%%%%%%%%%%%%%%%%%%%%%%%%%%%%%%%%%%%%

\section{Introduction}
%%%%%%%%%%%%%%%%%%%%%%%%%%%%%%%%%%%%%%%%%%%
Neutrinos play a crucial role in the physics of supernova explosions \cite{ja96} and in the 
early evolution of their compact stellar remnants \cite{bu86,ja95}. During the collapse of the 
pre-supernova core, a large number of neutrinos is produced by electron capture process.
The mean free path $\lambda$ of these neutrinos decreases as the radius of the newly formed 
neutron star shrinks from about $100$ km to about $10$ km, becoming smaller than the stellar 
radius when density reaches a critical value ({\it neutrino trapping density}). Under these
conditions neutrinos are {\it trapped} in the star. Neutrino trapping has a strong influence 
on the overall {\it stiffness} of the dense matter Equation of State (EoS) \cite{bo96,pr97}, being 
the physical conditions of the hot and lepton-rich newborn neutron star substantially different 
from those of the cold and deleptonized neutron star. 

The scattering of neutrinos on neutrons is mediated by the neutral current of the electroweak
interaction. In the non-relativistic limit for neutrons, the mean free path of a neutrino with 
initial energy ${\rm E}_\nu$ is given by~\cite{iwa82}
\begin{eqnarray} 
\lambda^{-1}({\rm E}_\nu,T) = \frac{G_F^2}{16 \pi^2} \int d{\bf k}_3  
\Bigg( c_V^2 (1+\cos{\theta})~{\cal S}^{(0)}(q,T) \nonumber \\
+ c_A^2 (3-\cos{\theta})~{\cal S}^{(1)}(q,T) \Bigg)~,
\label{eq1}
\end{eqnarray}
where $T$ is the temperature, $G_F$ is the Fermi constant, $c_V$
($c_A$) the vector (axial) coupling constant, $k_1=(E_\nu,{\bf k}_1)$ and $k_3$ are
the initial and final neutrino four-momenta, $q = k_1-k_3$ the 
transferred four-momentum, and $\cos\theta=\hat{\bf k}_1\cdot\hat{\bf
k}_3$. 
In the following, we impose the average energy $E_\nu=3T$~\cite{red98}.
The dynamical structure factors ${\cal S}^{(S)}(q,T)$
describe the response of neutron matter to excitations induced
by neutrinos, and they contain the relevant information on the
medium (cf Eq.~\ref{eq:structure}).
The vector (axial) part of the neutral current gives
rise to density (spin-density) fluctuations, corresponding to
the $S=0$ ($S=1$) spin channel.

In this paper, we report on calculations of the mean 
free path of neutrinos in pure neutron matter under various conditions of
density and temperature. A microcopic framework based on the Brueckner--Hartree--Fock (BHF)
approximation of the Brueckner--Bethe--Goldstone (BBG) theory is employed to describe consistently
both the EoS and the dynamical structure factors of neutron matter including
finite temperature effects. 
%We show that the lack of full consistency in the inclusion of
%thermal effects leads to an underestimation of the neutrino mean free path of about
%30-50\% for T=80 MeV.

The paper is organized in the following way. A brief review of the BHF
approximation at zero temperature and its extension to the finite temperature case is 
presented in Sec.\ \ref{sec:eos}. The Landau 
parameters $F_0, F_1, G_0$ and $G_1$ are calculated in Sec.\ \ref{sec:landau}. Section \ref{sec:spf} is 
devoted to the calculation of the dynamical structure factors and the neutrino mean free path. Finally, the main 
conclusions of this work are drawn in Sec.\ \ref{sec:conclusions}.

%%%%%%%%%%%%%%%%%%%%%%%%%%%%%%%%%%%%%%%%%%%%%%%%%%%%%%%%%%%%%%%%%%%%%%%%%%%%%%%%%%%%%%%%%%%

\section{EoS at finite temperature}
\label{sec:eos}
%%%%%%%%%%%%%%%%%%%%%%%%%%%%%%%%%%%%%%
\begin{figure}
\centering
\includegraphics[scale=0.2]{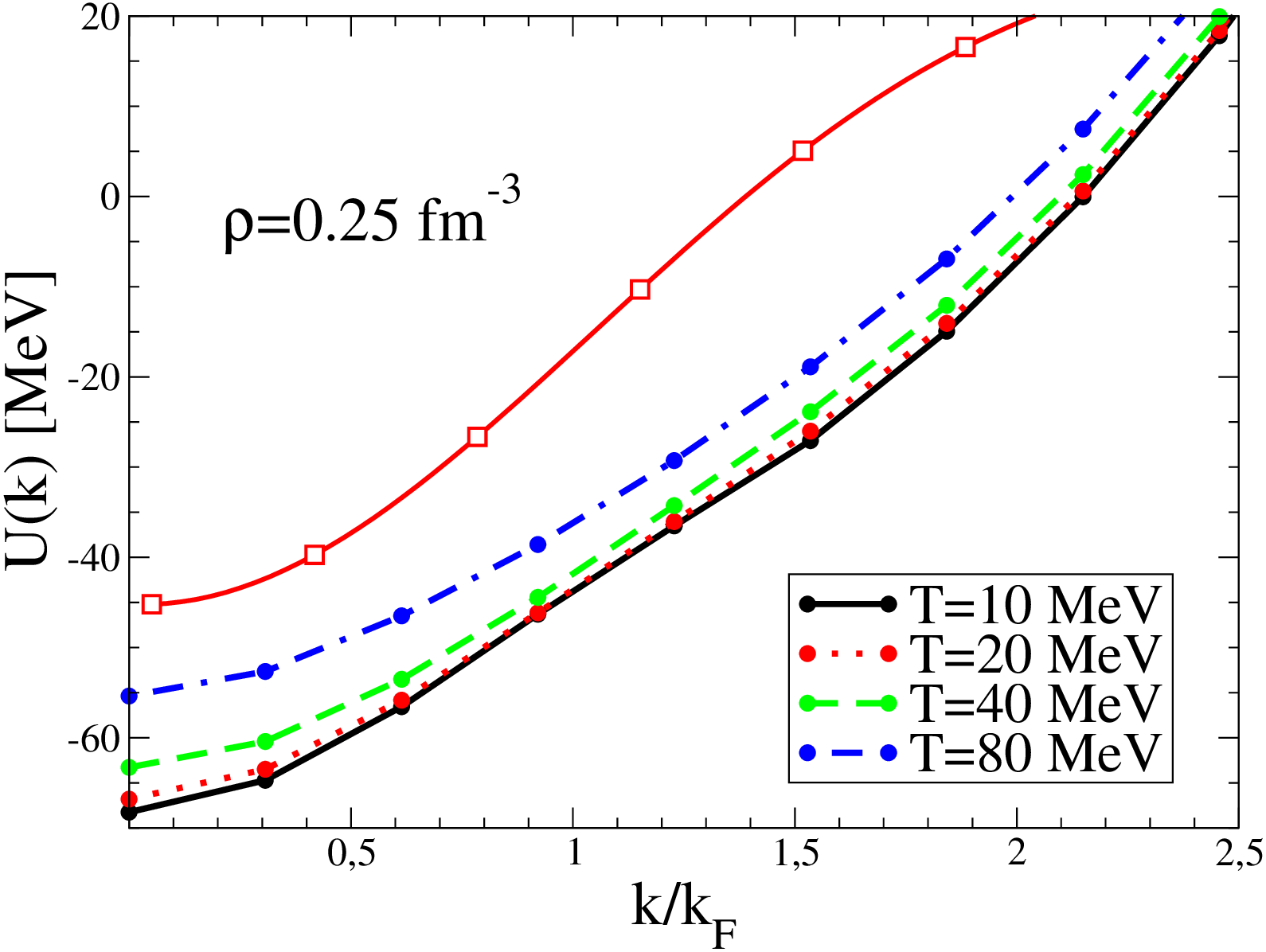}
\caption{Mean field for several temperatures as a function of the momentum k in
units of $k_F$. The result of Brueckner-Hartree-Fock calculation is represented with filled 
circles while the mean field given by the D1P Gogny
effective interaction is represented with empty squares.}
\label{fig1}
\end{figure}
%%%%%%%%%%%%%%%%%%%%%%%%%%%%%%%%%%%%%%

Our many-body scheme is based on the BHF approximation of the 
BBG theory. It starts with the construction of the 
neutron-neutron $G$-matrix, which describes in an effective way the interaction between 
two neutrons in the presence of a surrounding medium. It is formally obtained by solving 
the well known Bethe--Goldstone equation, written schematically as
\begin{equation}
G(\omega) = V + V \frac{Q}{\omega-E_1-E_2+ i\eta}G(\omega)  \ .
\label{eq:gmatrix}
\end{equation}
In the above expression $V$ denotes the bare interaction, $Q$ is the Pauli operator which 
allows only intermediate states compatible with the Pauli principle, and $\omega$, the so-called 
starting energy, corresponds to the sum of nonrelativistic single-particle energies of the 
interacting neutrons. The single-particle energy $E$ is given by 
\begin{equation}
E(k)=\frac{\hbar^2k^2}{2m}+U(k) \ ,
\label{eq:spe}
\end{equation}
where the single-particle potential $U(k)$ represents the mean field ``felt'' by the neutron 
due to its interaction with the other neutrons of the medium. In the BHF approximation $U(k)$ is 
given by
\begin{equation}
U(k) ={\mathrm Re} \sum_{k'}n(k') \left\langle \vec{k}\vec{k'}\right | G(\omega=E(k)+E(k')) \left | 
\vec{k}\vec{k'} \right\rangle _{\cal A}  
\label{eq:upot}
\end{equation}
where
\begin{equation}
n(k)=
\left\{\begin{array}{ll}
1, \makebox{if $k \leq k_{F}$} \\
0, \makebox{otherwise}  \\
\end{array} \right. \
\label{eq:ocnumb}
\end{equation}
is the corresponding occupation number and the matrix elements are properly antisymmetrized. The
resulting single-particle potential is shown toghether with the
corresponding one for the D1P Gogny \cite{far99} effective interaction in Fig.\ \ref{fig1} 
for $\rho=0.25$ fm$^{-3}$ ($k_F=1.95$ fm$^{-3}$). We note here that the so-called 
continuous prescription has been adopted for the 
single-particle potential when solving the Bethe--Goldstone equation. As shown by the authors of Refs. 
\cite{so98,ba00}, the contribution to the energy per particle from three-body clusters is diminished 
in this prescription. We note also that the present calculations have been carried out by using the 
nucleon-nucleon part of the recent realistic baryon-baryon interaction (model NSC97e) constructed by
the Nijmegen group \cite{st99}. The total energy per particle, $E/A$, is easily calculated once a 
self-consistent solution of Eqs.\
(\ref{eq:gmatrix})--({\ref{eq:upot}) is achieved
\begin{equation}
\frac{E}{A} = \frac{1}{A}\sum_{k} n(k)
\left(\frac{\hbar^2k^2}{2m}
+\frac{1}{2}U(k)\right) \ .
\label{eq:ea}
\end{equation}

%%%%%%%%%%%%%%%%%%%%%%%%%%%%%%%%%%%%%%
\begin{figure}
\centering
\includegraphics[scale=0.33]{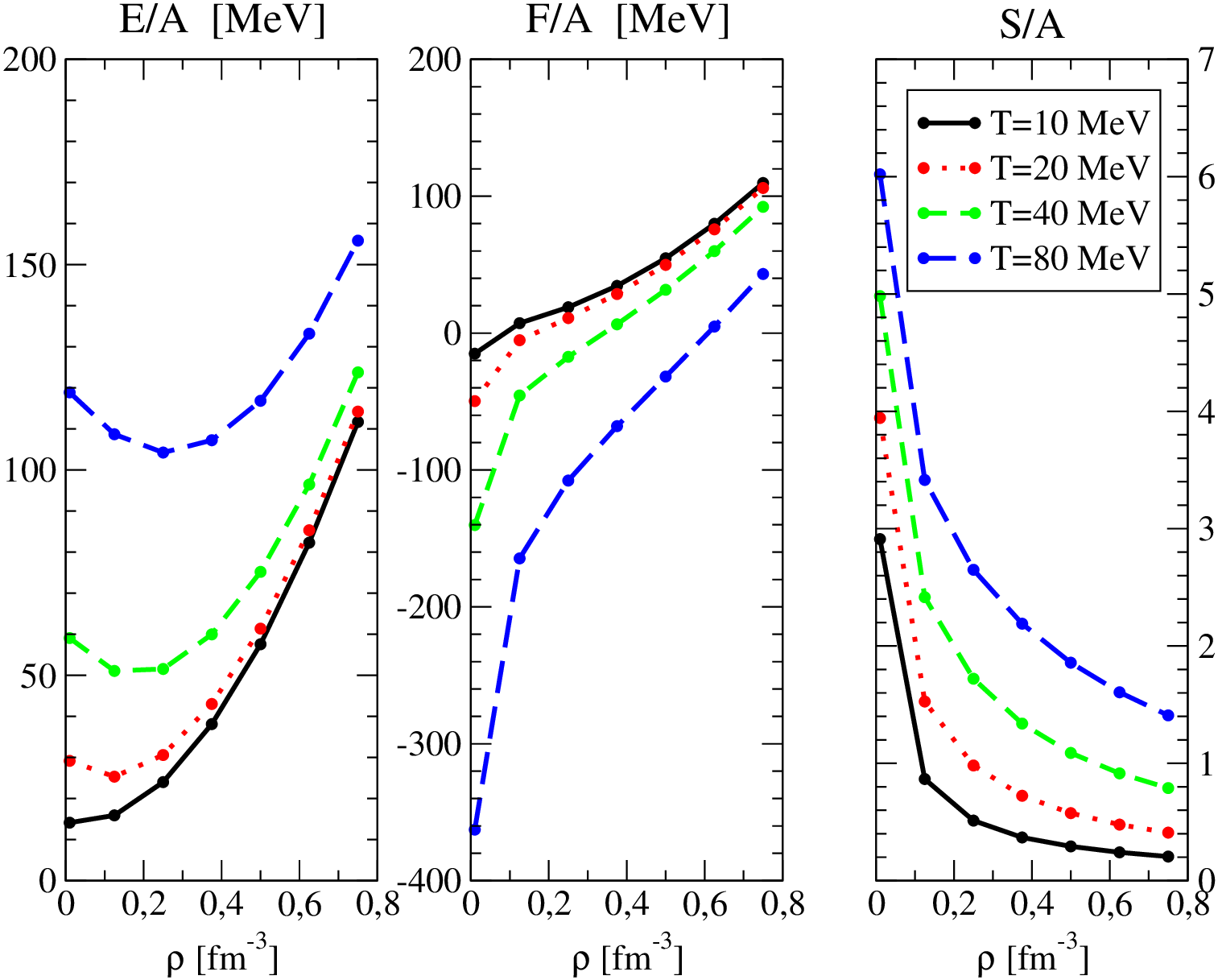}
\caption{Energy per particle, free energy per particle and entropy per
particle as a function of the number density for several temperatures.}
\label{fig1bis}
\end{figure}
%%%%%%%%%%%%%%%%%%%%%%%%%%%%%%%%%%%%%%

The many-body problem at finite temperature has been considered by several authors within different
approches, such as the finite temperature Green's function method \cite{fe71}, the thermo-field
method \cite{he95}, or the Bloch--De Domicis (BD) diagrammatic expansion \cite{bl58}. The latter, 
developed soon after the Brueckner theory, represents the ``natural'' extension to finite temperature 
of the BBG expansion, to which it leads in the zero temperature limit. Baldo and Ferreira \cite{ba99} 
showed that the dominant terms in the BD expansion were those that correspond to the zero temperature 
BBG diagrams, where the temperature is introduced only through the Fermi-Dirac distribution
\begin{equation}
f(k,T)=\frac{1}{1+\exp\left([E(k,T)-\mu_n(T)]/T\right)} \ .
\label{eq:fd}
\end{equation}

Therefore, at the BHF level, finite temperature effects can be introduced in a very good 
approximation just replacing in the Bethe--Goldstone equation: (i) the zero temperature Pauli operator
$Q=(1-n_1)(1-n_2)$ by the corresponding finite temperature one $Q(T)=(1-f_1)(1-f_2)$, and (ii) the 
single-particle energies $E(k)$ by the temperature dependent ones $E(k,T)$ obtained from Eqs.\ 
(\ref{eq:spe}-\ref{eq:upot}) when $n(k)$ is replaced by $f(k,T)$.

In this case, however, the self-consistent process implies that together with the Bethe--Goldstone 
equation and the single-particle potential, the chemical potential of the neutron, $\mu_n(T)$, must 
be extracted at each step of the iterative process from the normalization
condition
\begin{equation}
\rho=\sum_{k} f(k,T) \ .
\label{eq:cp}
\end{equation}
This is an implicit equation which can be solved numerically by e.g., the Newton--Rapson method. Note 
that, now, also the Bethe--Goldstone equation and single-particle potential depend implicitly on 
the chemical potential.
\begin{figure}
\centering
\includegraphics[scale=0.3]{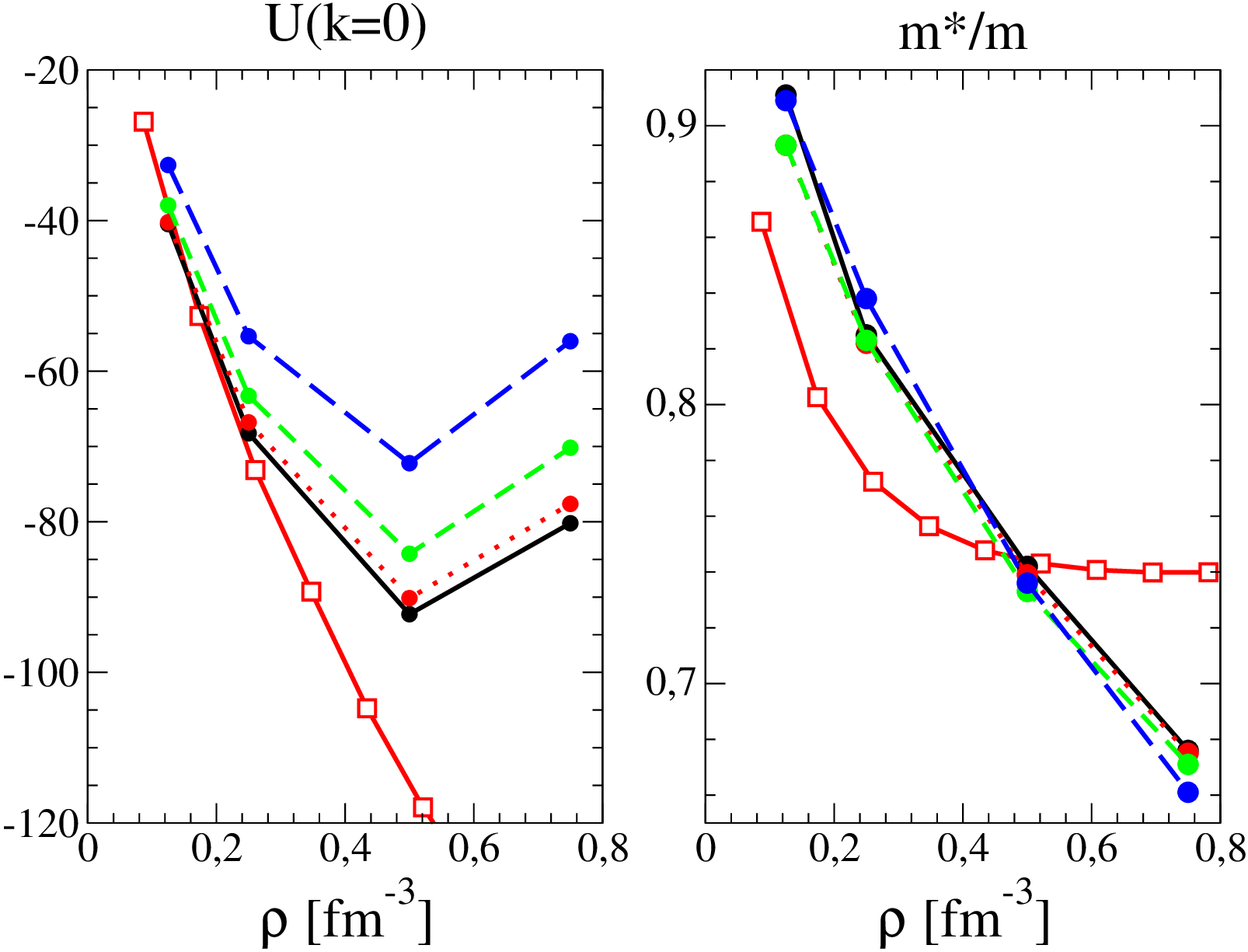}
\caption{Quadratic approximation of the single-particle energy: $U(k=0)$ is the constant
mean field and $m^*$ is the effective mass. See Fig.~\ref{fig1} for the legend.}
\label{fig2}
\end{figure}

Once a self-consistent solution is obtained the total free energy per particle is determined by
\begin{equation}
\frac{F}{A}=\frac{E}{A}-T\frac{S}{A} \ ,
\label{eq:free_ener}
\end{equation}
where $E/A$ is evaluated from Eq.\ (\ref{eq:ea}) replacing $n(k)$ by $f(k,T)$ and the
total entropy per particle, $S/A$, is calculated through the expression
\begin{eqnarray}
\frac{S}{A}=-\frac{1}{A}\sum_{k}\left[f(k,T)\ln(f(k,T)) \right.\nonumber \\
\left. +\left(1-f(k,T)\right)\ln(1-f(k,T))\right] \ .
\label{eq:entrop}
\end{eqnarray}
Results for the quantities $E/A$, $F/A$ and $S/A$ are shown in Fig.\ \ref{fig1bis} 
for several densities and
temperatures.

The k-momentum dependence of the single-particle energy is usually approximated 
by a quadratic function which consist in a constant term, the mean field, 
and a squared momentum dependent term, related to the neutron effective mass $m^*$, 
reading
\be
E^{\rm eff}(k) \equiv \frac{\hbar^2 \k^2}{2 m^*}+U(k=0)
\label{eq:quadratic}
\ee
where
\be
\frac{\hbar}{m^*}=\frac{\hbar}{m}+\frac{1}{\k}\frac{\partial U(k)}{\partial \k}
|_{\k=\k_F}
\ee
We finish this section by showing in Fig.~\ref{fig2} the mean field and the effective mass 
as a function of the density, for several temperatures. As the temperature increases, the value of the 
mean field also increases.  One can remark that the effective mass is nearly independent of 
the temperature up to T=80 MeV.
%As the effective mass is related to the non-locality of the
%potential, it is related to the width of the bare interaction, hence, it reveals
%that the masses of the mesons are not affected by the temperature.

%%%%%%%%%%%%%%%%%%%%%%%%%%%%%%%%%%%%%%%%%%%%%%%%%%%%%%%%%%%%%%%%%%%%%%%%%%%%%%%%%%%%%%%%%%%

\section{Landau parameters $F_0$, $F_1$, $G_0$ and $G_1$}
\label{sec:landau}

%%%%%%%%%%%%%%%%%%%%%%%%%%%%%%%%%%%%%%
\begin{figure}
\centering
\includegraphics[scale=0.3]{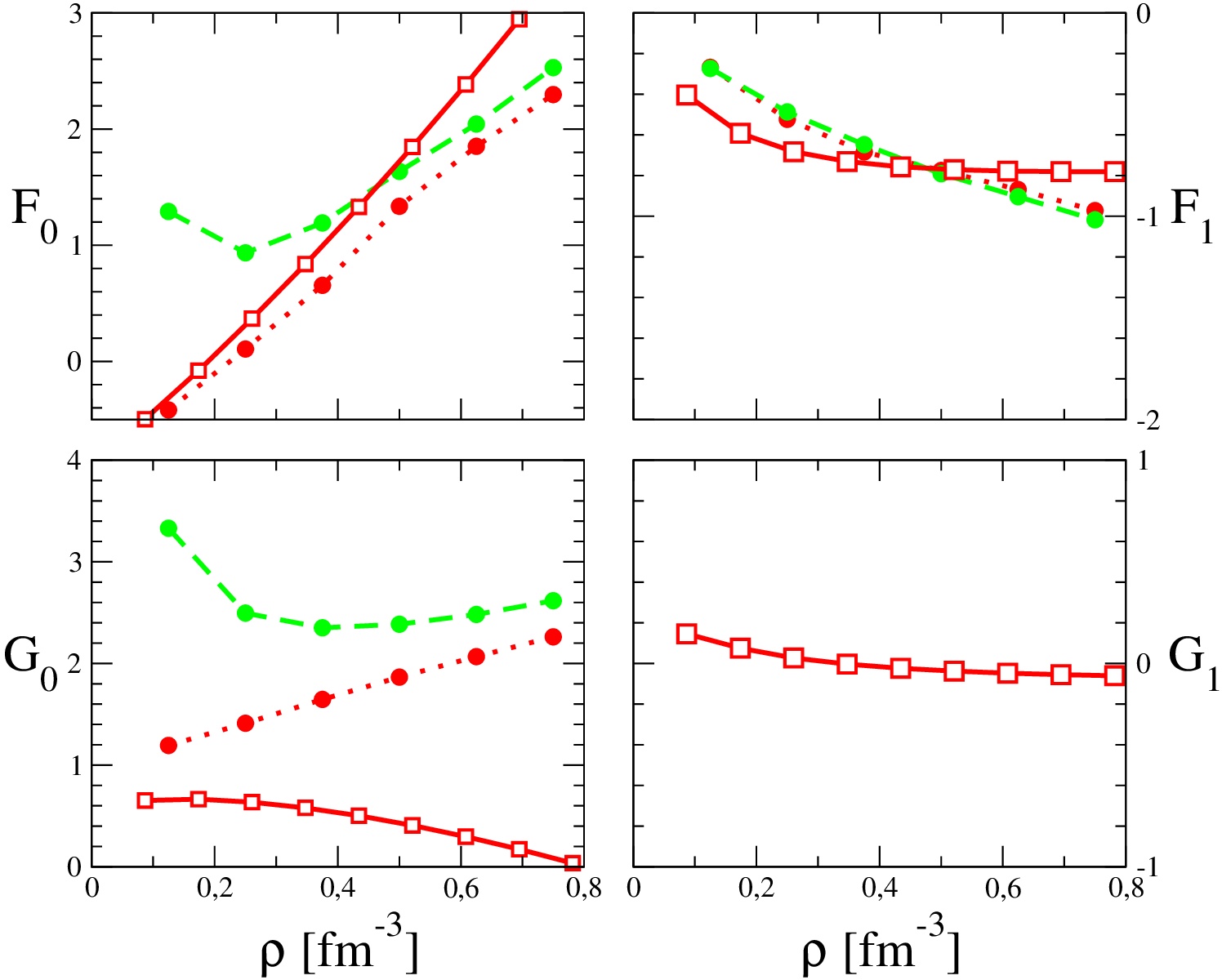}
\caption{Landau parameters deduced for T=10 MeV (dotted line)
and 80 MeV (dashed line) using the Nijmegen potential. It is compared with the 
one deduced from the D1P Gogny interaction (empty squares).} 
\label{fig4}
\end{figure}
%%%%%%%%%%%%%%%%%%%%%%%%%%%%%%%%%%%%%%

The Landau parameters $F_0$, $G_0$ are related to the incompressibility modulus
$K$ and the magnetic susceptibility $\chi$ respectivly through 
\begin{equation}
K=9\rho \frac{\partial^2 F/V }{\partial\rho^2}=9\rho\frac{1+F_0}{N_0} \ ,
\end{equation} 
\begin{equation}
\chi^{-1}=\frac{1}{(\mu\rho)^2}\frac{\partial^2 F/V}{\partial\rho_s^2}=\frac{1+G_0}{\chi_F} \ ,
\end{equation} 
where $F/V=(F/A)\rho$ is the free energy density, $N_0=m^* k_F/\pi^2\hbar^2$ is
the density of states, $\chi_F$ is the magnetic susceptibility of a free Fermi gas and
$\rho_s=\rho_\uparrow-\rho_\downarrow$ is the spin asymmetry density (i.e., the difference
in the densities of neutrons with spin up and spin down). The parameter
$F_1$ is deduced from the effective mass according to the relation
\begin{equation}
\frac{m^*}{m}=1+\frac{F_1}{3} \ .
\end{equation}
There is no simple relation for the Landau parameter $G_1$ that can be deduced
from thermodynamical properties or general relations like the forward scattering sum rule. 
Gogny effective interaction predicts that $G_1$ is close to zero at all densities 
(cf Fig.~\ref{fig4}), hence, we will assume in the following that $G_1$=0.

Finally, the Landau parameters we obtain in pure neutron matter are shown
on Fig.~\ref{fig4}. The absolute values of $F_0$ and $G_0$ increase 
with the temperature while $F_1$ is nearly constant. The main differences between 
the Landau parameters calculated with the BHF approximation and those calculated 
with the Gogny effective interaction, are present in the spin fluctuation channel. 
In the latter case, for high densities, the Landau parameter $G_0$ falls to $-1$ 
because of a ferromagnetic instability \cite{mar02} while $G_0$ increase with 
density for the BHF calculation, since no such an instability is found in this case 
\cite{vid02a}.

\section{Dynamical Structure Factors and Neutrino mean free path}
\label{sec:spf}

We obtain the dynamical
structure factors ${\cal S}^{(S)}$ from the imaginary part of
the response function $\chi^{(S)}(q,T)$ in pure neutron matter 
as~\cite{nav99}
\begin{equation}
{\cal S}^{(S)}(\q,\q_0,T) = -\frac{1}{\pi}\frac{1}{1-\exp(-\q_0/T)} 
\,\,\,\Im {\rm m} \,\,\chi^{(S)}(\q,\q_0,T) \, .
\label{eq:structure}
\end{equation}
In the Brueckner--Hartree--Fock (BHF) approximation the response function is given by
\begin{equation}
\Im {\rm m} \,\,\chi^{(S)}_{\rm BHF}(\q,\q_0,T) = -\frac{{m^*}^2 T}{4\pi \q}\ln
\left(\frac{1+\e^{(A+\q_0/2)/T}}{1+\e^{(A-\q_0/2)/T}} \right) \, ,
\label{eq:response}
\end{equation}
where $A=\mu_n-m^*/2(\q_0/\q)^2-\q^2/8m^*$.
This expression is valid for both positive and negative transferred energies
$\q_0$. The real and imaginary part of the response function have been obtain
within the quadratic approximation for the single-particle energy (see Eq.\ (\ref{eq:quadratic})).
The influence of the in-medium interaction appears through the effective mass $m^*$ and the 
chemical potential $\mu_n$, which introduce, besides a density dependence, an additional 
temperature dependence in the response function.

We have also included the long range correlations within the BHF+RPA scheme.
The particle-hole interaction is approached by a Landau form, 
containing only $l=0$ multipole (RPA $l=0$), or $l=0, 1$ multipoles 
(RPA $l=0,1$) \cite{mar01}.
We stress that in all cases the particle-hole interaction has 
been consistently obtained from the particle-particle interaction 
used to describe the EoS.

%%%%%%%%%%%%%%%%%%%%%%%%%%%%%%%%%%%%%%
\begin{figure}
\centering
\includegraphics[scale=0.25]{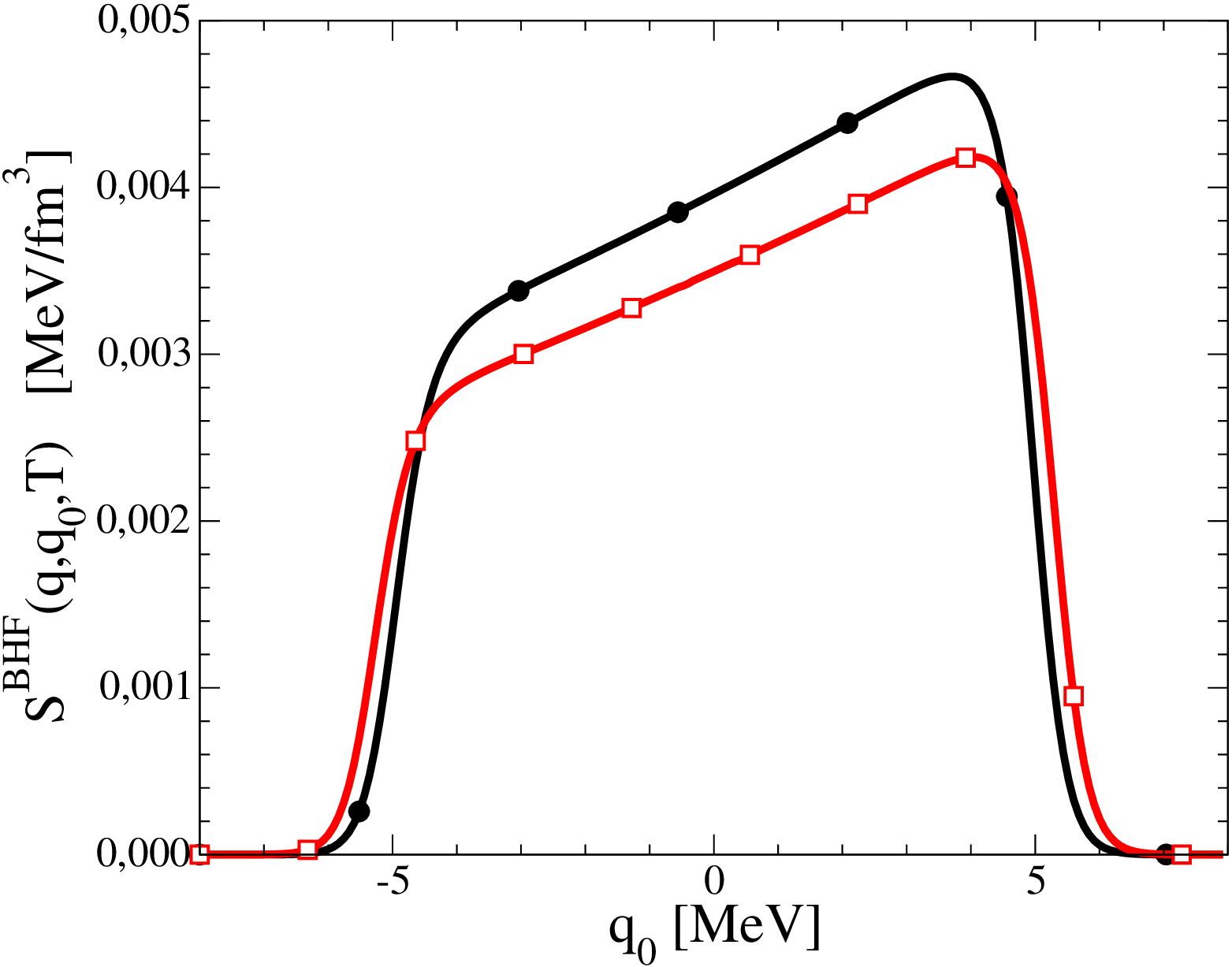}
\caption{Dynamical structure factors at the BHF approximation as a 
function of the transfered energy ${\rm q}_0$, and for $\rho=0.25$ fm, 
T=10 MeV and ${\rm q}=10$ MeV. }
\label{fig5a}
\end{figure}
%%%%%%%%%%%%%%%%%%%%%%%%%%%%%%%%%%%%%%
%%%%%%%%%%%%%%%%%%%%%%%%%%%%%%%%%%%%%%
\begin{figure}
\centering
\includegraphics[scale=0.33]{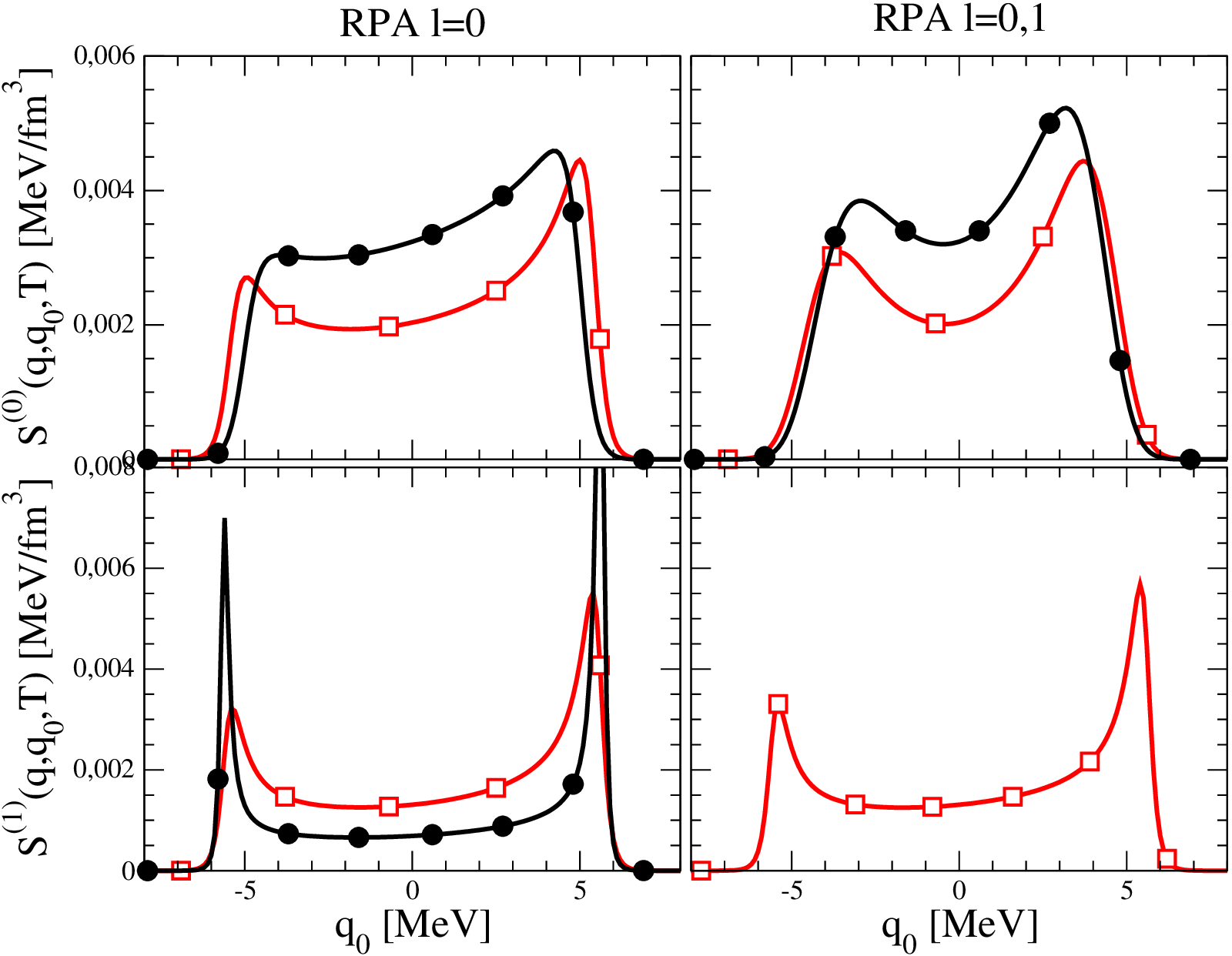}
\caption{Dynamical structure factors, for the channel $S=0$
  and $S=1$, as a function of the transfered energy ${\rm q}_0$, and for 
$\rho$=0.25 fm$^{-3}$, T=10 MeV and ${\rm q}$=10 MeV. }
\label{fig5b}
\end{figure}
%%%%%%%%%%%%%%%%%%%%%%%%%%%%%%%%%%%%%%
The dynamical structure factors, within the BHF and BHF+RPA scheme, are displayed
in Fig.~\ref{fig5a} and Fig.~\ref{fig5b} as a function of the transferred 
energy ${\rm q}_0$ and for $\rho$=0.25 fm$^{-3}$, T=10 MeV and ${\rm q}=10$ 
MeV. In Fig.~\ref{fig5b} we show the dynamical structure factors for 
the density ($S=0$) and spin-density ($S=1$) channels.
We compare the response function obtained with the Brueckner input (filled 
circle) to the result with the D1P Gogny effective interaction (empty square).
The two columns correspond to two truncations in the BHF+RPA calculation which 
consist to include only the $l=0$ Landau parameters or the $l=0,1$ Landau 
parameters.
As expected from the values of the Landau parameters (cf Fig.~\ref{fig4}),
the main differences between the approaches are present in the spin channel ($S=1$).
The spin sound is weakly present for the D1P Gogny effective interaction while it seems very 
important for the Brueckner calculation. 
%These differences reflect the difference between the Landau parameters $G_0$ (see Fig.~\ref{fig4}).
The pronounced zero sound in the spin-density channel at zero temperature is damped when
the temperature increases, and these differences disappear at high temperature.

%%%%%%%%%%%%%%%%%%%%%%%%%%%%%%%%%%%%%%
\begin{figure}[htb]
\centering
\includegraphics[scale=0.3]{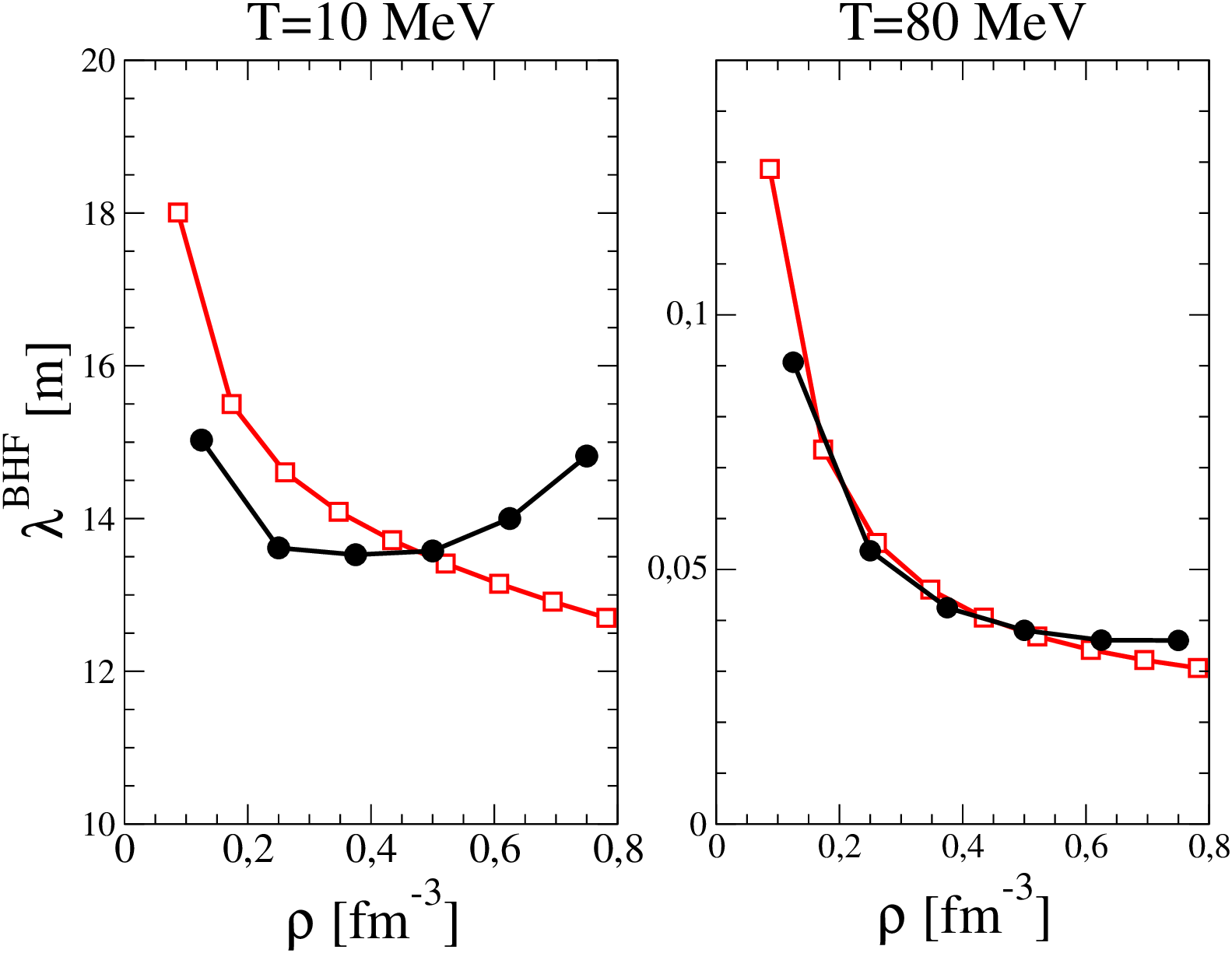}
\caption{The neutrino mean free path deduced from BHF response function as a function of 
the density for T=10 MeV (left) and T=80 MeV (right). 
The energy of the incoming neutrino is $E_\nu=3T$.}
\label{fig6}
\end{figure}
%%%%%%%%%%%%%%%%%%%%%%%%%%%%%%%%%%%%%%

The density and temperature dependence of the neutrino mean free path is
shown in Fig.~\ref{fig6}. We have also performed a comparison with
Gogny prediction. The behaviour of the mean free path follows approximatively
the expected law in $T^{-2}$~\cite{red98}.
At low temperature, the discrepancies between the two interactions are 
essentially due to the density dependance of the effective mass 
(cf Eq.~\ref{eq:response}).
Moreover, as the effective mass is independent of the temperature, a 
calculation of the in-medium interaction at finite temperature is not
necessary.
Hence, within BHF scheme, the mean free path is sensible to the temperature 
essentially through the explicit T-dependance of the response function and
the effect due to the finite temperature EoS is negligable.
At high temperature, the 2 interactions gives very similar results because the
effects of the medium are reduced ($E_\nu=3T$).

The effects of the residual interaction, i.e., of RPA correlations, can be seen 
in Fig.~\ref{fig7}, where we have represented the ratio 
$R=\lambda^{\rm BHF+RPA l=0,1}/\lambda^{\rm BHF}$ of the neutrino mean
free path within BHF+RPA $l=0,1$ to BHF as a function of the density. 
The solid line stands for T=10 MeV and the dotted line for T=80 MeV.
The ratio $R$ with Nijmegen potential is about 8 at high density while it is
only 1 with Gogny interaction. 
This effect have also been shown for zero temperature EoS with three-body 
forces~\cite{lom02}.
This is due to the presence of spin instabilities at
high density in the case of Gogny interaction~\cite{mar01,mar03}. These
instabilities lead to the divergence of the dynamical form factor~\cite{mar02}, 
hence the neutrino mean free path goes to zero.
This illustrate the sensibility of the neutrino mean free path with the onset of 
instabilities. Here, it reduces the mean free path by a factor 8 at high density.

%%%%%%%%%%%%%%%%%%%%%%%%%%%%%%%%%%%%%%
\begin{figure}[htb]
\centering
\includegraphics[scale=0.25]{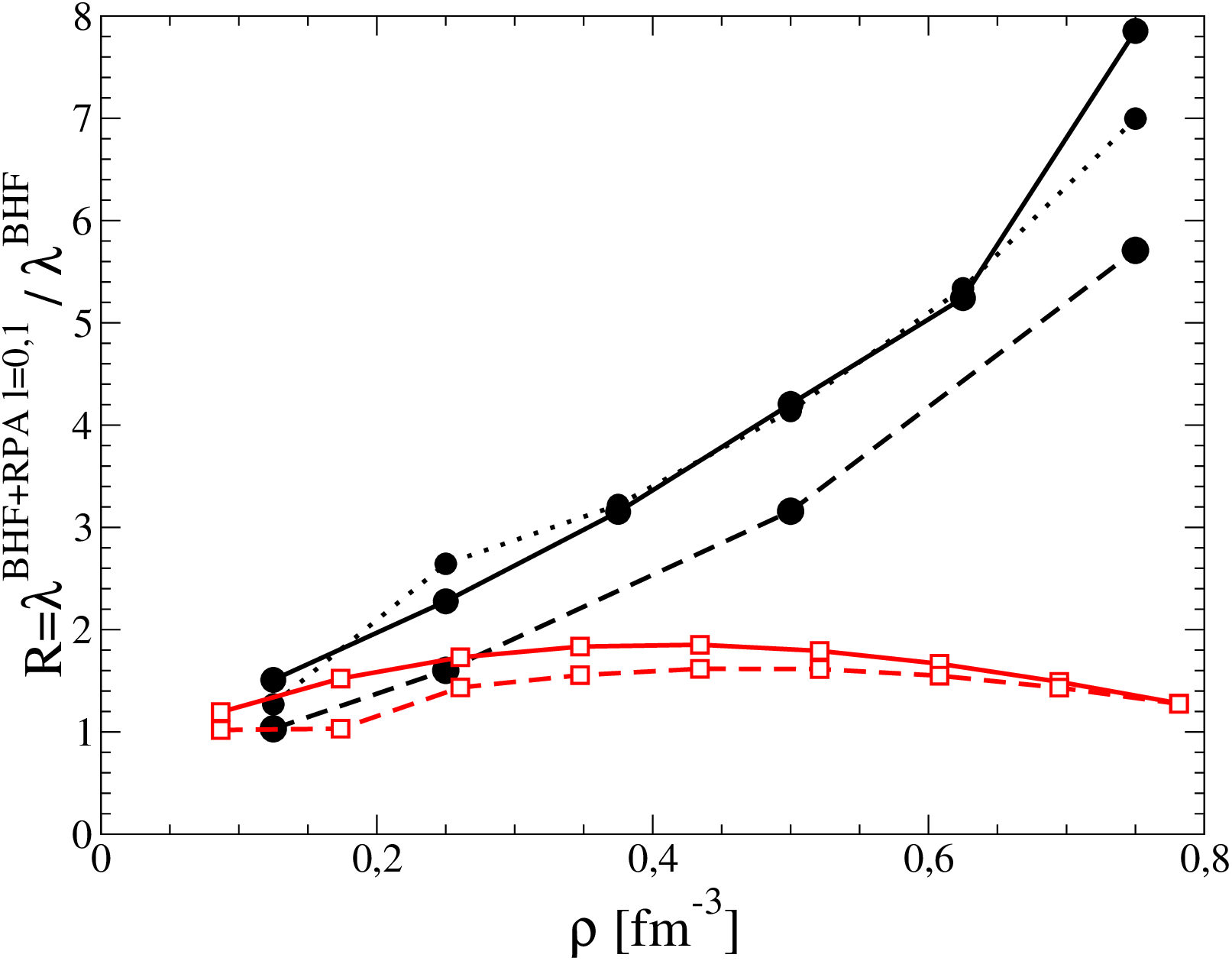}
\caption{Effects of BHF+RPA $l=0,1$ for Brueckner (filled circle) and Gogny interactions 
(empty square). The solid line stands for a consistent calculation at
T=10 MeV, the dashed line is for T=80 MeV but with the EoS at T=10 MeV (consistency broken)
and the dotted line is for T=80 MeV (consistent).
The energy of the incoming neutrino is $E_\nu=3T$.}
\label{fig7}
\end{figure}
%%%%%%%%%%%%%%%%%%%%%%%%%%%%%%%%%%%%%%

In order to understand the interplay between the explicit T-dependance of the 
response function (cf Eq.~\ref{eq:response}) and the Landau parameters calculated
at finite temperature, we have broken the consistency of the approach. 
In Fig.~\ref{fig7}, the solid line stands for a consistent calculation at
T=10 MeV while the dashed line stands for the calculation of the neutrino
mean free path using low 
temperature Landau parameters (T=10 MeV) but calculating the dynamical response
function at T=80 MeV. 
Hence, by comparing the solid and the dashed line calculated with the same 
particle-hole residual interaction, we see that the increase of temperature 
reduces the effect of the correlations. As an illustration, the increase of 
the temperature suppress the collective modes. The increase of temperature
induce also the increase of the incoming neutrino energy ($E_\nu=3T$). 
The lack of consistency between finite temperature EoS and the dynamical response
function leads to an underestimation of the neutrino mean free path of about 30-50\%.
The effect of restauring the temperature consistency for T=80 MeV is illustrated 
by comparing the dashed line and the dotted line (consistent calculation for T=80 MeV). 
As it has been shown on Fig.~\ref{fig4}, the Landau parameters $F_0$ and $G_0$ increase with the
temperature. The correlations become more important and the ratio $R$ increase.
Finally, the temperature induce two opposite effects: it increases the particle-hole
interaction, but it decreases the effects of these correlations in the calculation
of the dynamical structure factors. These two effects tend to compensate each other 
and the ratio $R$, at low temperature, is close to the ratio at high temperature.

%%%%%%%%%%%%%%%%%%%%%%%%%%%%%%%%%%%%%%%%%%%%%%%%%%%%%%%%%%%%%%%%%%%%%%%%%%%%%%%%%%%%%%

\section{Conclusions}
\label{sec:conclusions}

The purpose of this article is to study consistently the effects of the 
temperature to compute the neutrino mean free path in dense neutron matter. 
Both the EoS and the response function have been computed, including the BHF+RPA 
correlations. We have shown that the effective mass is nearly independent of the
temperature.

We have shown the interplay between the effects of temperature for the calculation
of the EoS (the mean field and residual interaction deduced from it)
and the explicit temperature dependance of the dynamical structure factor.
The first one increases the correlations with the temperature while the second
one decreases it. Finally, these two opposite effects approximatively
compensate each other and the ratio $R$ is
nearly constant for the range of temperature 10-80 MeV.

The temperature consistency does not change the general behaviour of the neutrino mean
free path. The lack of consistency leads to an underestimation of the neutrino mean free path
of about 30-50\% for T=80 MeV.

\section*{Acknowledgements}

The authors are very grateful to professor J. Navarro
for useful discussions and comments. One of the authors (J.M.) wishes to acknowledge 
the hospitality and support of the Istituto Nazionale di Fisica Nucleare, 
sezione di Pisa (Italy).

\appendix

\end{document}